\documentclass[a4paper,11pt]{article}
\usepackage{jheppub} 
\usepackage{hyperref}
\usepackage[T1]{fontenc} 

\title{\boldmath Transition magnetic moment of Majorana neutrinos in the $\mu\nu$SSM}

\author[a,b]{Hai-Bin Zhang,}
\author[a,b,c,d]{Tai-Fu Feng,}
\author[a]{Zhao-Feng Ge}
\author[a,c,d]{and Shu-Min Zhao}

\affiliation[a]{Department of Physics, Hebei University, \\ Baoding, 071002,  China}
\affiliation[b]{Department of Physics, Dalian University of Technology, \\
Dalian, 116024, China}
\affiliation[c]{Institute of theoretical Physics, Chinese Academy of Sciences, \\
Beijing 100190, China}
\affiliation[d]{The Key Laboratory of Mathematics-Mechanization (KLMM), \\
Beijing 100190, China}

\emailAdd{hbzhang@mail.dlut.edu.cn}
\emailAdd{fengtf@hbu.edu.cn}
\emailAdd{algezhaofeng@126.com}
\emailAdd{smzhao@hbu.edu.cn}

\abstract{The nonzero vacuum expectative values of sneutrinos induce spontaneously R-parity and lepton number violation, and generate three tiny Majorana neutrino masses through the seesaw mechanism in the $\mu\nu$SSM, which is one of Supersymmetric extensions beyond Standard Model. Applying effective Lagrangian method, we study the transition magnetic moment of Majorana neutrinos in the model here. Under the constraints from neutrino oscillations, we consider the two possibilities on the neutrino mass spectrum with normal or inverted ordering.}


\begin{document}
\maketitle
\flushbottom

\section{Introduction}
In the Standard Model (SM), neutrinos have only weak interaction and are massless. However, the observations of neutrino oscillations (see refs.~\cite{neu-b1,neu-b2,neu-b3,neu-b4}) imply that neutrinos have very small masses and are mixed. Therefore, the SM must be extended to account for the neutrino masses and mixing. In any extensions of the SM, the transition magnetic moments of neutrinos are induced through electroweak radiative corrections. In addition, the transition magnetic moments of neutrinos can generate important effects, especially in astrophysical environments, where neutrinos propagate for long distances in magnetic fields
both in matter and in vacuum~\cite{Broggini}.

Applying effective Lagrangian method and on-shell scheme, here we analyze the radiative contributions from one-loop diagrams to the transition magnetic moment of Majorana neutrinos in the ``$\mu$ from $\nu$ Supersymmetric Standard Model'' ($\mu\nu$SSM)~\cite{mnSSM,mnSSM1,mnSSM2}, where nonzero vacuum expectative values (VEVs) of sneutrinos lead to R-parity and lepton number violations. The $\mu\nu$SSM could generate three tiny massive Majorana neutrinos at the tree level through the mixing with the neutralinos and right-handed neutrinos~\cite{mnSSM,mnSSM1,mnSSM2,meu-m,meu-m1,meu-m2,meu-m3,Zhang2}. Especially, the $\mu$ problem~\cite{m-problem} of the Minimal Supersymmetric Standard Model (MSSM)~\cite{MSSM,MSSM1,MSSM2,MSSM3,MSSM4} had been solved in the $\mu\nu$SSM, through the R-parity breaking couplings ${\lambda _i}\hat \nu _i^c\hat H_d^a\hat H_u^b$ in the superpotential. The $\mu$ term is generated spontaneously via the nonzero VEVs of right-handed sneutrinos, $\mu  = {\lambda _i}\left\langle {\tilde \nu _i^c} \right\rangle$, when the electroweak symmetry is broken (EWSB).

Three flavor neutrinos $\nu_{e,\mu,\tau}$ are mixed into the mass eigenstates $\nu_{1,2,3}$ during their flight, and the mixing is described by the Pontecorvo-Maki-Nakagawa-Sakata unitary matrix
$U_{_{PMNS}}$~\cite{neutrino-oscillations,neutrino-oscillations1}.  Through the several recently updated neutrino oscillation experiments T2K~\cite{T2K}, MINOS~\cite{MINOS}, DOUBLE CHOOZ~\cite{DOUBLE-CHOOZ}, RENO~\cite{RENO} and especially Daya Bay~\cite{DAYA-BAY}, the neutrino mixing angle $\theta_{13}$ is now precisely known. One can give fitted value for $\theta_{13}$ as~\cite{Palazzo}
\begin{eqnarray}
\sin^2 2\theta_{13}=0.090\pm 0.009\,.
\label{neu-oscillations1}
\end{eqnarray}
The other experimental observations of the neutrino oscillation parameters in $U_{_{PMNS}}$ show that~\cite{PDG}
\begin{eqnarray}
&&\;\:\Delta m_{\odot}^2 =7.58_{-0.26}^{+0.22}\times 10^{-5} {\rm eV}^2\,,\qquad
\sin^2\theta_{12} =0.306_{-0.015}^{+0.018}\,,\nonumber\\
&&|\Delta m_{A}^2| =2.35_{-0.09}^{+0.12}\times 10^{-3} {\rm eV}^2\,,\qquad \sin^2\theta_{23}=0.42_{-0.03}^{+0.08}\,.
\label{neu-oscillations2}
\end{eqnarray}
In our numerical analysis, we use the neutrino experimental data presented in eq.~(\ref{neu-oscillations1}) and eq.~(\ref{neu-oscillations2}) to constrain the input parameters in the model here. Assuming neutrino mass spectrum with normal or inverted ordering, we analyse the transition magnetic moment of Majorana neutrinos in the $\mu\nu$SSM.

The outline of the paper is as follow. In section~\ref{sec2}, we outline the $\mu\nu$SSM by introducing its superpotential and the general soft SUSY-breaking terms. In section~\ref{sec3}, we will give formulae of the transition magnetic moment of Majorana neutrinos in the $\mu\nu$SSM, applying effective Lagrangian method and on-shell scheme. The results of our numerical study are given in section~\ref{sec4}. And the conclusions are left for section~\ref{sec5}. Finally, the relative mass matrices, couplings and form factors are collected in appendices~\ref{appendix-mass}--\ref{appendix-factor}.

\section{The $\mu\nu$SSM}
\label{sec2}

Besides the superfields of the MSSM, the $\mu\nu$SSM introduces three singlet right-handed neutrino superfields $\hat{\nu}_i^c\;(i=1,\;2,\;3)$. The corresponding superpotential of the $\mu\nu$SSM is given by~\cite{mnSSM}
\begin{eqnarray}
W =&&{\epsilon _{ab}} \Big( {Y_{u_{ij}}}\hat H_u^b\hat Q_i^a\hat u_j^c + {Y_{d_{ij}}}\hat H_d^a\hat Q_i^b\hat d_j^c
+ {Y_{e_{ij}}}\hat H_d^a\hat L_i^b\hat e_j^c + {Y_{\nu _{ij}}}\hat H_u^b\hat L_i^a\hat \nu _j^c \Big)  \nonumber\\
&&- {\epsilon _{ab}}{\lambda _i}\hat \nu _i^c\hat H_d^a\hat H_u^b + \frac{1}{3}{\kappa _{ijk}}\hat \nu _i^c\hat \nu _j^c\hat \nu _k^c \, ,
\end{eqnarray}
where $\hat H_d^T = \Big( {\hat H_d^0,\hat H_d^ - } \Big)$, $\hat H_u^T = \Big( {\hat H_u^ + ,\hat H_u^0} \Big)$, $\hat Q_i^T = \Big( {{{\hat u}_i},{{\hat d}_i}} \Big)$, $\hat L_i^T = \Big( {{{\hat \nu}_i},{{\hat e}_i}} \Big)$ are $SU(2)$ doublet superfields, and $\hat d_i^c$, $\hat u_i^c$ and $\hat e_i^c$ represent the singlet down-type quark, up-type quark
and charged lepton superfields, respectively. In addition, $Y_{u,d,\nu,e}$, $\lambda$, $\kappa$ respectively are dimensionless matrices, a vector, a totally symmetric tensor. And $a,b$ are SU(2) indices with antisymmetric tensor $\epsilon_{12}=-\epsilon_{21}=1$. The summation convention is implied on repeated indices in this paper. In the superpotential, the first three terms are the same as the MSSM. Next two terms can generate the effective bilinear terms $\epsilon _{ab} \varepsilon_i \hat H_u^b\hat L_i^a$, $\epsilon _{ab} \mu \hat H_d^a\hat H_u^b$,  and $\varepsilon_i= Y_{\nu _{ij}} \left\langle {\tilde \nu _j^c} \right\rangle$, $\mu  = {\lambda _i}\left\langle {\tilde \nu _i^c} \right\rangle$,  once the electroweak symmetry is broken. The last term generates the effective Majorana masses for neutrinos at the electroweak scale. And the last two terms explicitly violate lepton number and R-parity.

In the framework of supergravity mediated supersymmetry breaking, the general soft SUSY-breaking terms in the $\mu\nu$SSM are given as
\begin{eqnarray}
- \mathcal{L}_{soft}=&&m_{{{\tilde Q}_{ij}}}^{\rm{2}}\tilde Q{_i^{a\ast}}\tilde Q_j^a
+ m_{\tilde u_{ij}^c}^{\rm{2}}\tilde u{_i^{c\ast}}\tilde u_j^c + m_{\tilde d_{ij}^c}^2\tilde d{_i^{c\ast}}\tilde d_j^c
+ m_{{{\tilde L}_{ij}}}^2\tilde L_i^{a\ast}\tilde L_j^a  \nonumber\\
&&+ \; m_{\tilde e_{ij}^c}^2\tilde e{_i^{c\ast}}\tilde e_j^c + m_{{H_d}}^{\rm{2}} H_d^{a\ast} H_d^a
+ m_{{H_u}}^2H{_u^{a\ast}}H_u^a + m_{\tilde \nu_{ij}^c}^2\tilde \nu{_i^{c\ast}}\tilde \nu_j^c \nonumber\\
&&+ \; \epsilon_{ab} \Big[{{({A_u}{Y_u})}_{ij}}H_u^b\tilde Q_i^a\tilde u_j^c
+ {{({A_d}{Y_d})}_{ij}}H_d^a\tilde Q_i^b\tilde d_j^c + {{({A_e}{Y_e})}_{ij}}H_d^a\tilde L_i^b\tilde e_j^c + \textrm{H.c.} \Big] \nonumber\\
&&+ \; \Big[ {\epsilon _{ab}}{{({A_\nu}{Y_\nu})}_{ij}}H_u^b\tilde L_i^a\tilde \nu_j^c
- {\epsilon _{ab}}{{({A_\lambda }\lambda )}_i}\tilde \nu_i^c H_d^a H_u^b
+ \frac{1}{3}{{({A_\kappa }\kappa )}_{ijk}}\tilde \nu_i^c\tilde \nu_j^c\tilde \nu_k^c + \textrm{H.c.} \Big] \nonumber\\
&&- \; \frac{1}{2}\Big({M_3}{{\tilde \lambda }_3}{{\tilde \lambda }_3}
+ {M_2}{{\tilde \lambda }_2}{{\tilde \lambda }_2} + {M_1}{{\tilde \lambda }_1}{{\tilde \lambda }_1} + \textrm{H.c.} \Big)\,.
\end{eqnarray}
Here, the first two lines consist of mass squared terms of squarks, sleptons and Higgses. The next two lines contain the trilinear scalar couplings. In the last lines, $M_3$, $M_2$ and $M_1$ denote Majorana masses corresponding to $SU(3)$, $SU(2)$ and $U(1)$ gauginos $\hat{\lambda}_3$, $\hat{\lambda}_2$ and $\hat{\lambda}_1$, respectively.
In addition to the terms from $\mathcal{L}_{soft}$, the tree-level scalar potential receives the usual D and F term contributions~\cite{mnSSM1}.

Once the electroweak symmetry is spontaneously broken, the neutral scalars develop in general the following VEVs:
\begin{eqnarray}
\langle H_d^0 \rangle = \upsilon_d \,, \qquad \langle H_u^0 \rangle = \upsilon_u \,, \qquad
\langle \tilde \nu_i \rangle = \upsilon_{\nu_i} \,, \qquad \langle \tilde \nu_i^c \rangle = \upsilon_{\nu_i^c} \,.
\end{eqnarray}
Thus one can define neutral scalars as usual
\begin{eqnarray}
&&H_d^0=\frac{h_d + i P_d}{\sqrt{2}} + \upsilon_d, \qquad\; \tilde \nu_i = \frac{(\tilde \nu_i)^\Re + i (\tilde \nu_i)^\Im}{\sqrt{2}} + \upsilon_{\nu_i},  \nonumber\\
&&H_u^0=\frac{h_u + i P_u}{\sqrt{2}} + \upsilon_u, \qquad \tilde \nu_i^c = \frac{(\tilde \nu_i^c)^\Re + i (\tilde \nu_i^c)^\Im}{\sqrt{2}} + \upsilon_{\nu_i^c}.
\end{eqnarray}
And one can define
\begin{eqnarray}
\tan\beta=\frac{\upsilon_u}{\sqrt{\upsilon_d^2+\upsilon_{\nu_i}\upsilon_{\nu_i}}}.
\end{eqnarray}

For simplicity, we will assume that all parameters in the potential are real in the model. After EWSB, the charge scalar mass matrix $M_{S^{\pm}}^2$, neutral fermion mass matrix $M_n$ and charged fermion mass matrix $M_c$ are given in appendix~\ref{appendix-mass}. The charged scalar mass matrix $M_{S^{\pm}}^2$ contains massless unphysical Goldstone bosons $G^{\pm}$, which can be written as~\cite{Zhang,Zhanghb,Zhang1}
\begin{eqnarray}
G^{\pm} = {1 \over \sqrt{\upsilon_u^2 + \upsilon_d^2+\upsilon_{\nu_i}\upsilon_{\nu_i}}} \Big(\upsilon_d H_d^{\pm} - \upsilon_u {H_u^{\pm}}+\upsilon_{\nu_i}\tilde e_{L_i}^{\pm}\Big)\,.
\end{eqnarray}
In the physical gauge, the Goldstone bosons $G^{\pm}$ are eaten by $W$-boson, and disappear from the Lagrangian. And the mass squared of $W$-boson is
\begin{eqnarray}
m_W^2= \frac{e^2}{2 s_{_W}^2} \Big(\upsilon_u^2 + \upsilon_d^2+\upsilon_{\nu_i}\upsilon_{\nu_i} \Big),
\end{eqnarray}
where $e$ is the electromagnetic coupling constant, $s_{_W}=\sin\theta_{_W}$ (and $c_{_W}=\cos\theta_{_W}$ below) with the Weinberg angle $\theta_{_W}$, respectively.

In the $\mu\nu$SSM, left-handed neutrinos mix with the neutralinos and right-handed neutrinos. Via the seesaw mechanism \cite{mnSSM}, the effective light neutrino mass matrix is in general given as
\begin{eqnarray}
{m_{eff}} =  -  m.{M^{ - 1}} .{m^T},
\end{eqnarray}
where the concrete expressions for the mass matrices $M$ and $m$ are given in appendix~\ref{appendix-mass}. Diagonalized the effective neutrino mass matrix ${m_{eff}}$, we can obtain three light neutrino masses.

\section{Neutrino magnetic moment}
\label{sec3}

The magnetic dipole moment (MDM) and electric dipole moment (EDM) of the Dirac fermion (including charged lepton and neutrino etc) can be actually be written as the operators
\begin{eqnarray}
&&\mathcal{L}_{MDM}=\frac{1}{2} \mu_{ij} \bar{\psi}_i \sigma^{\mu\nu} \psi_j F_{\mu\nu},\nonumber\\
&&\:\mathcal{L}_{EDM}=\frac{i}{2} \epsilon_{ij} \bar{\psi}_i \sigma^{\mu\nu} \gamma_5 \psi_j F_{\mu\nu},
\label{MEDM}
\end{eqnarray}
where $\sigma^{\mu\nu}=\frac{i}{2}[\gamma^\mu,\gamma^\nu]$, $F_{\mu\nu}$ is the electromagnetic field strength, $\psi_{i,j}$ denote the four-component Dirac fermions which are on-shell, $\mu_{ij}$ and $\epsilon_{ij}$ are Dirac diagonal ($i=j$) or transition
($i\neq j$) MDM and EDM between states $\psi_{i}$ and $\psi_{j}$, respectively.

In fact, it convenient to get the contributions from loop diagrams to fermion diagonal or transition MDM and EDM in terms of the effective Lagrangian method, if the masses $m_{_V}$ of internal lines are much heavier than the external fermion mass $m_f$. Since ${/\!\!\! p}=m_f \ll m_{_V}$ for on-shell fermion and ${/\!\!\! k}\rightarrow 0 \ll m_{_V}$ for photon, we can expand the amplitude of corresponding triangle diagrams according to the external momenta of fermion and photon. After matching between the effective theory and the full theory, we get all high dimension operators together with their coefficients. It is enough to retain only those dimension 6 operators in later calculations~\cite{Feng,Feng1,Feng2}:
\begin{eqnarray}
&&O_1^{L,R} = e \bar{\psi}_i {(i {/\!\!\!\! \mathcal{D}})}^3 P_{L,R} \psi_j , \nonumber\\
&&O_2^{L,R} = e \overline{(i \mathcal{D}_\mu {\psi}_i )} \gamma^\mu F\cdot \sigma P_{L,R} \psi_j, \nonumber\\
&&O_3^{L,R} = e \bar{\psi}_i  F\cdot \sigma \gamma^\mu P_{L,R} {(i \mathcal{D}_\mu {\psi}_j )}, \nonumber\\
&&O_4^{L,R} = e \bar{\psi}_i  (\partial^\mu F_{\mu\nu})  \gamma^\nu P_{L,R} \psi_j, \nonumber\\
&&O_5^{L,R} = e m_{{\psi}_i} \bar{\psi}_i  {(i {/\!\!\!\! \mathcal{D}})}^2 P_{L,R} \psi_j, \nonumber\\
&&O_6^{L,R} = e m_{{\psi}_i} \bar{\psi}_i F\cdot \sigma P_{L,R} \psi_j,
\label{operators}
\end{eqnarray}
where $\mathcal{D}_\mu=\partial^\mu+ieA_\mu$,  $P_L=\frac{1}{2}{(1 - {\gamma _5})}$, $P_R=\frac{1}{2}{(1 + {\gamma _5})}$ and $m_{{\psi}_i}$ is the mass of fermion ${{\psi}_i}$.

Certainly, all dimension 6 operators in eq.~(\ref{operators}) induce the effective couplings among photons and fermions. The effective vertices with one external photon are written as
\begin{eqnarray}
&&O_1^{L,R} = ie \{((p+k)^2+p^2)\gamma_\rho+({/\!\!\! p}+{/\!\!\! k})\gamma_\rho{/\!\!\! p}\} P_{L,R}, \nonumber\\
&&O_2^{L,R} = ie ({/\!\!\! p}+{/\!\!\! k})[{/\!\!\! k}, \gamma_\rho] P_{L,R}, \nonumber\\
&&O_3^{L,R} = ie [{/\!\!\! k}, \gamma_\rho] {/\!\!\! p} P_{L,R}, \nonumber\\
&&O_4^{L,R} = ie  (k^2\gamma_\rho-{/\!\!\! k}k_\rho) P_{L,R}, \nonumber\\
&&O_5^{L,R} = ie m_{{\psi}_i} \{({/\!\!\! p}+{/\!\!\! k})\gamma_\rho+\gamma_\rho {/\!\!\! p}\}  P_{L,R}, \nonumber\\
&&O_6^{L,R} = ie m_{{\psi}_i} [{/\!\!\!k}, \gamma_\rho]P_{L,R}.
\end{eqnarray}

If the full theory is invariant under the combined transformation of charge conjugation, parity and time reversal (CPT), the induced effective theory preserves the symmetry after the heavy freedoms are integrated out. The fact implies the Wilson coefficients of the operators $O_{2,3,6}^{L,R}$ satisfying the relations~\cite{Feng}
\begin{eqnarray}
C_3^{L,R}=C_2^{R,L\ast}, \qquad  C_6^{L}=C_6^{R\ast}\,,
\end{eqnarray}
where $C_I^{L,R}$ ($I = 1\cdots6$) represent the Wilson coefficients of the corresponding operators $O_I^{L,R}$ in the effective Lagrangian. After applying the equations of motion to the external fermions, we find that the concerned terms in the effective Lagrangian are transformed into
\begin{eqnarray}
&&\quad\; C_2^{R}O_2^R + C_2^{L}O_2^L +  C_2^{L\ast}O_3^R + C_2^{R\ast}O_3^L + C_6^{R}O_6^R + C_6^{R\ast}O_6^L \nonumber\\
&&\Rightarrow (C_2^{R} + \frac{m_{{\psi}_j}}{m_{{\psi}_i}}C_2^{L\ast} + C_6^{R})O_6^R + (C_2^{R\ast} + \frac{m_{{\psi}_j}}{m_{{\psi}_i}}C_2^{L} + C_6^{R\ast})O_6^L \nonumber\\
&&=e m_{{\psi}_i} \Re(C_2^{R} + \frac{m_{{\psi}_j}}{m_{{\psi}_i}}C_2^{L\ast} + C_6^{R})\bar{\psi}_i \sigma^{\mu\nu} \psi_j  F_{\mu\nu} \nonumber\\
&&\quad +\: ie m_{{\psi}_i}\Im(C_2^{R} + \frac{m_{{\psi}_j}}{m_{{\psi}_i}}C_2^{L\ast} + C_6^{R})\bar{\psi}_i \sigma^{\mu\nu} \gamma_5 \psi_j   F_{\mu\nu},
\label{CtoRI}
\end{eqnarray}
where, $\Re(\cdots)$ and $\Im(\cdots)$ denote the operation to take the real and imaginary part of a complex number, respectively. Matching between eq.~(\ref{MEDM}) and eq.~(\ref{CtoRI}), we can obtain
\begin{eqnarray}
&&\mu_{ij} = 4m_e m_{{\psi}_i} \Re(C_2^{R} + \frac{m_{{\psi}_j}}{m_{{\psi}_i}}C_2^{L\ast} + C_6^{R}) \mu_{\rm{B}},\nonumber\\
&&\:\epsilon_{ij} = 4m_e m_{{\psi}_i} \Im(C_2^{R} + \frac{m_{{\psi}_j}}{m_{{\psi}_i}}C_2^{L\ast} + C_6^{R}) \mu_{\rm{B}},
\end{eqnarray}
where $\mu_{\rm{B}}=e/2 m_e$ and $m_e$ is the electron mass. In other words, the MDM and EDM of the Dirac fermions are respectively proportional to real and imaginary part of the effective coupling $C_2^{R} + \frac{m_{{\psi}_j}}{m_{{\psi}_i}}C_2^{L\ast} + C_6^{R}$.

As a neutral fermion, the mass eigenstate of neutrino, may be not just Dirac field but also Majorana field. The Majorana neutrino coincides with its antiparticle. The four degrees of freedom of a Dirac neutrino (two helicities and two particle-antiparticle) are reduced to two (two helicities) by the Majorana constraint. The electromagnetic properties of the Majorana neutrino are possibly reduced, because the Majorana neutrino just has half the degrees of freedom of the Dirac neutrino. Through the general description of the electromagnetic form factors of Dirac and Majorana neutrinos in ref.~\cite{Broggini}, here we can get the MDM and EDM for Majorana neutrinos
\begin{eqnarray}
&&\mu_{ij}^M = \mu_{ij}^D -\mu_{ji}^D,\qquad\epsilon_{ij}^M = \epsilon_{ij}^D -\epsilon_{ji}^D,
\label{Majorana-MDM}
\end{eqnarray}
with
\begin{eqnarray}
&&\mu_{ij}^D = 4m_e m_{\nu_i} \Re(C_2^{R} + \frac{m_{\nu_j}}{m_{\nu_i}}C_2^{L\ast} + C_6^{R}) \mu_{\rm{B}},\nonumber\\
&&\:\epsilon_{ij}^D = 4m_e m_{\nu_i} \Im(C_2^{R} + \frac{m_{\nu_j}}{m_{\nu_i}}C_2^{L\ast} + C_6^{R}) \mu_{\rm{B}},
\end{eqnarray}
where $\nu_{i,j}$  denote Majorana neutrinos. In eq.~(\ref{Majorana-MDM}), the first terms $\mu_{ij}^D$ and $\epsilon_{ij}^D$ denote Dirac-neutrino-like terms for the MDM and EDM of Majorana neutrinos, the second terms $-\mu_{ji}^D$ and $-\epsilon_{ji}^D$ denote Dirac-antineutrino-like terms for the MDM and EDM of Majorana neutrinos, respectively. One also can find that $\mu_{ij}^M$ and $\epsilon_{ij}^M$ are antisymmetric. So, Majorana neutrinos don't have diagonal MDM and EDM, but can have transition MDM and EDM.

\begin{figure}
\setlength{\unitlength}{1mm}
\centering
\begin{minipage}[c]{0.43\textwidth}
\includegraphics[width=2.5in]{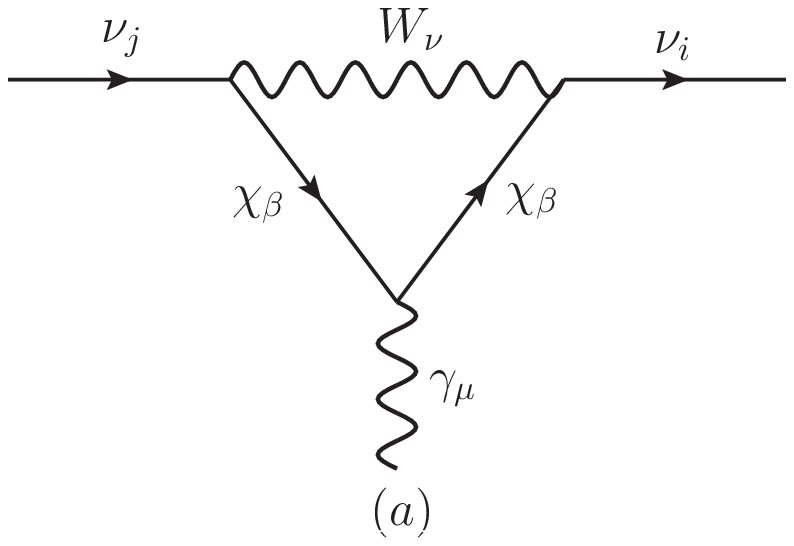}
\end{minipage}%
\begin{minipage}[c]{0.43\textwidth}
\includegraphics[width=2.5in]{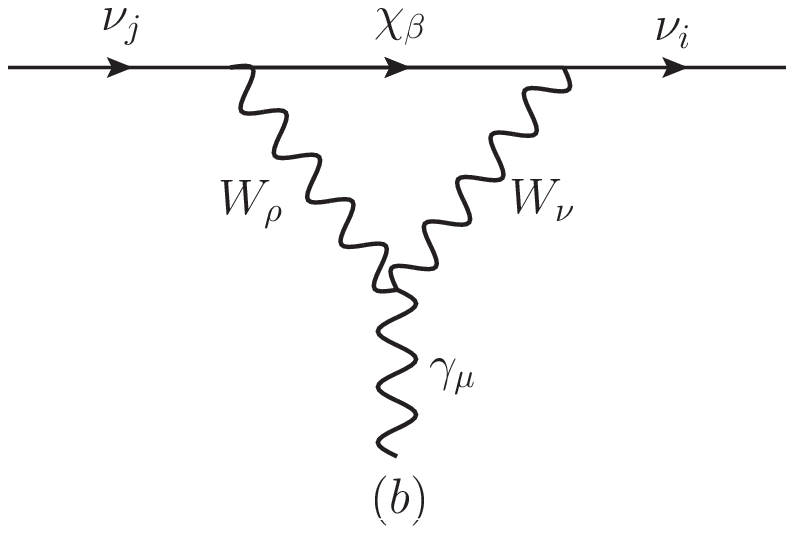}
\end{minipage}
\begin{minipage}[c]{0.43\textwidth}
\includegraphics[width=2.5in]{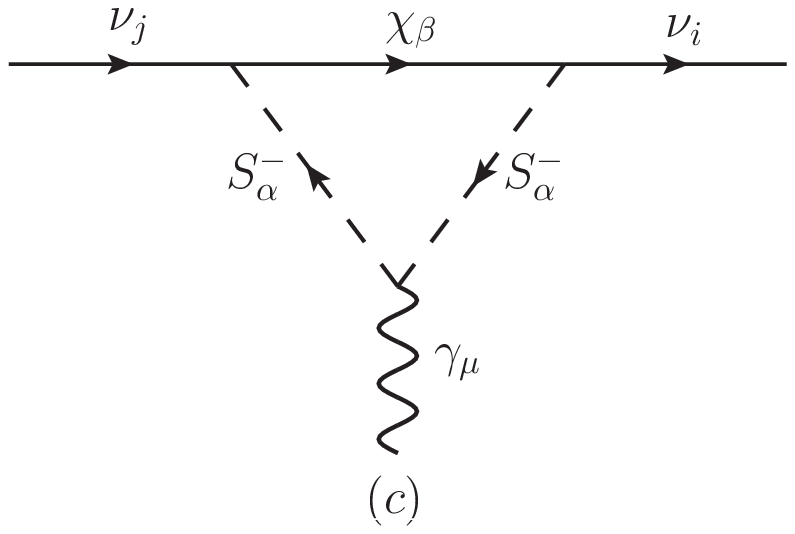}
\end{minipage}%
\begin{minipage}[c]{0.43\textwidth}
\includegraphics[width=2.5in]{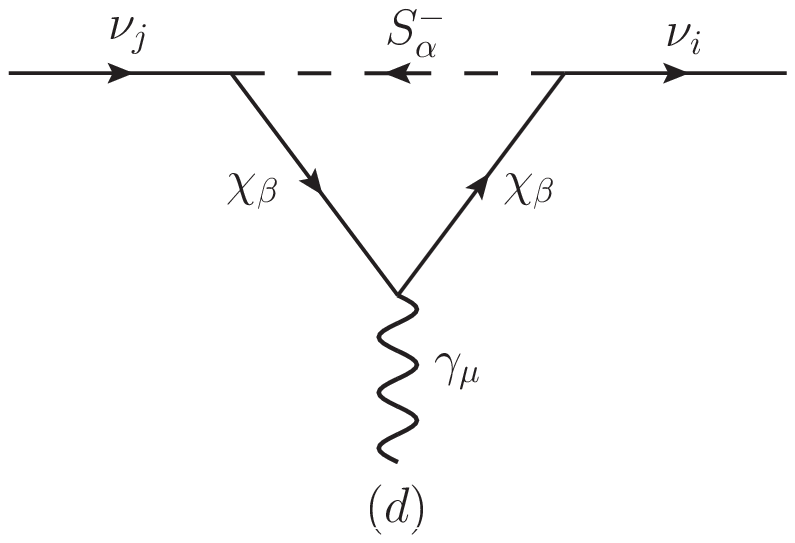}
\end{minipage}
\caption[]{One-loop diagrams contributing to transition magnetic moment of Majorana neutrinos in the $\mu\nu$SSM, where (a) and (b) represent the charged fermions $\chi_\beta$ and $W$-boson loop contributions, (c) and (d) represent the charged fermions $\chi_\beta$ and charged scalars $S_\alpha ^-$ loop contributions.}
\label{feynman}
\end{figure}

Within framework of the $\mu\nu$SSM, we can obtain three light massive Majorana neutrinos through the seesaw mechanism. Next we will analyse the dominant one-loop diagrams contributing to the transition magnetic moment of the Majorana neutrinos in the $\mu\nu$SSM, which are depicted by figure~\ref{feynman}. Then the above Wilson coefficients can be written as a sum of four terms,
\begin{eqnarray}
C_{2,6}^{L,R}=C_{2,6}^{{L,R}(a)}+C_{2,6}^{{L,R}(b)}+C_{2,6}^{{L,R}(c)}+C_{2,6}^{{L,R}(d)}.
\end{eqnarray}
The terms $C_{2,6}^{{L,R}(a,b)}$ represent the charged fermions and $W$-boson loop contributions as
\begin{eqnarray}
&&C_2^{R(a)}= \frac{1}{2{m_W^2}} C_R^{W {\chi _\beta} \bar \chi _{7+i} ^ \circ } C_R^{W \chi _{7+j} ^ \circ {{\bar \chi }_\beta}} \Big[  {I_1}({x_{\chi _\beta },{x_{W }}}) - {I_4}({x_{\chi _\beta },{x_{W }}}) \Big], \nonumber\\
&&C_6^{R(a)}=  \frac{2{m_{\chi _\beta }}}{{m_W^2}{m_{\nu_i} }}  C_L^{W {\chi _\beta} \bar \chi _{7+i} ^ \circ } C_R^{W \chi _{7+j} ^ \circ {{\bar \chi }_\beta}} \Big[ {I_3}({x_{\chi _\beta },{x_{W }}}) -  {I_1}({x_{\chi _\beta },{x_{W }}}) \Big],  \nonumber\\
&&C_2^{R(b)}=  \frac{1}{2{m_W^2}} C_R^{W {\chi _\beta} \bar \chi _{7+i} ^ \circ } C_R^{W \chi _{7+j} ^ \circ {{\bar \chi }_\beta}} \Big[  {I_3}({x_{\chi _\beta },{x_{W }}}) +  {I_4}({x_{\chi _\beta },{x_{W }}})\Big], \nonumber\\
&&C_6^{R(b)}=  \frac{2{m_{\chi _\beta }}}{{m_W^2}{m_{\nu_i} }}   C_L^{W {\chi _\beta} \bar \chi _{7+i} ^ \circ } C_R^{W \chi _{7+j} ^ \circ {{\bar \chi }_\beta}} \Big[  -  {I_3}({x_{\chi _\beta },{x_{W }}})  \Big], \nonumber\\
&&C_{2,6}^{L(a,b)}=C_{2,6}^{R(a,b)}\mid _{L \leftrightarrow R},
\end{eqnarray}
where the concrete expressions for the coupling coefficients $C_{L,R}$ and form factors  $I_k\:(k=1,\cdots,4)$ can be found in appendix \ref{appendix-coupling} and appendix \ref{appendix-factor}, $x_a = {m_a^2}/{m_W^2}$ and $m_a$ is the mass for the corresponding particle, respectively.

Similarly, the charged fermions and charged scalars loop contributions $C_{2,6}^{{L,R}(c,d)}$ are
\begin{eqnarray}
&&C_2^{R(c)}=  \frac{1}{4{m_W^2}} C_R^{S_\alpha^{-\ast} {\chi _\beta} \bar \chi _{7+i} ^ \circ } C_L^{S_\alpha^- \chi _{7+j} ^ \circ {{\bar \chi }_\beta}} \Big[ {I_4}({x_{\chi _\beta },{x_{S_\alpha^- }}}) - {I_3}({x_{\chi _\beta },{x_{S_\alpha^- }}}) \Big], \nonumber\\
&&C_6^{R(c)}=  \frac{{m_{\chi _\beta }}}{2{m_W^2}{m_{\nu_i} }}  C_R^{S_\alpha^{-\ast} {\chi _\beta} \bar \chi _{7+i} ^ \circ } C_R^{S_\alpha^- \chi _{7+j} ^ \circ {{\bar \chi }_\beta}} \Big[ {I_3}({x_{\chi _\beta },{x_{S_\alpha^- }}}) - {I_1}({x_{\chi _\beta },{x_{S_\alpha^- }}})  \Big],  \nonumber\\
&&C_2^{R(d)}= \frac{1}{4{m_W^2}} C_R^{S_\alpha^{-\ast} {\chi _\beta} \bar \chi _{7+i} ^ \circ } C_L^{S_\alpha^- \chi _{7+j} ^ \circ {{\bar \chi }_\beta}} \Big[ 2{I_3}({x_{\chi _\beta },{x_{S_\alpha^- }}}) - {I_1}({x_{\chi _\beta },{x_{S_\alpha^- }}}) - {I_4}({x_{\chi _\beta },{x_{S_\alpha^- }}}) \Big], \nonumber\\
&&C_6^{R(d)}= \frac{{m_{\chi _\beta }}}{2{m_W^2}{m_{\nu_i} }}   C_R^{S_\alpha^{-\ast} {\chi _\beta} \bar \chi _{7+i} ^ \circ } C_R^{S_\alpha^- \chi _{7+j} ^ \circ {{\bar \chi }_\beta}} \Big[{I_1}({x_{\chi _\beta },{x_{S_\alpha^- }}})-{I_2}({x_{\chi _\beta },{x_{S_\alpha^-}}}) -   {I_3}({x_{\chi _\beta },{x_{S_\alpha^- }}}) \Big],  \nonumber\\
&& C_{2,6}^{L(c,d)}=C_{2,6}^{R(c,d)}\mid _{L \leftrightarrow R}.
\end{eqnarray}

\section{The numerical results}
\label{sec4}

\subsection{The parameter space}

It is well known that there are many free parameters in various supersymmetric extensions of the SM. In order to obtain a more transparent numerical results, we will do some assumptions on the concerned parameter space of the $\mu\nu{\rm SSM}$ in this paper. First, we adopt assumptions for some parameters:
\begin{eqnarray}
&&{\kappa _{ijk}} = \kappa {\delta _{ij}}{\delta _{jk}}, \quad\: {({A_\lambda }\lambda )_i} = {A_\lambda }\lambda, \qquad\: \lambda _i = \lambda, \qquad\; \upsilon_{\nu_i^c}=\upsilon_{\nu^c}, \nonumber\\
&&{Y_{{\nu _{ij}}}} = {Y_{{\nu _i}}}{\delta _{ij}}, \quad (A_\nu Y_\nu)_{ij}=a_{\nu _i}{\delta _{ij}}, \quad {Y_{{e_{ij}}}} = {Y_{{e_i}}}{\delta _{ij}}.
\end{eqnarray}
Restrained by the lepton masses, we have
\begin{eqnarray}
{Y_{{e_i}}} = \frac{{{m_{{l_i}}}}}{{{\upsilon_d}}},
\end{eqnarray}
where $m_{l_i}$ denote the charged lepton masses.

For soft breaking slepton mass matrices $m_{{{\tilde L},{\tilde e^c}}}^2$ and trilinear coupling matrices $({A_e}{Y_e})$, we will take into account the off-diagonal terms, which are defined as~\cite{sl-mix,sl-mix1,sl-mix2,sl-mix3,sl-mix4}
\begin{eqnarray}
&&\quad\;\,{m_{\tilde L}^2} = \left( {\begin{array}{*{20}{c}}
   1 & \delta_{12}^{LL} & \delta_{13}^{LL}  \\
   \delta_{12}^{LL} & 1 & \delta_{23}^{LL}  \\
   \delta_{13}^{LL} & \delta_{23}^{LL} & 1  \\
\end{array}} \right){m_{L}^2},\\
&&\quad\:{m_{\tilde e^c}^2} = \left( {\begin{array}{*{20}{c}}
   1 & \delta_{12}^{RR} & \delta_{13}^{RR}  \\
   \delta_{12}^{RR} & 1 & \delta_{23}^{RR}  \\
   \delta_{13}^{RR} & \delta_{23}^{RR} & 1  \\
\end{array}} \right){m_{E}^2},\\
&&({A_e}{Y_e}) = \left( {\begin{array}{*{20}{c}}
   m_{l_1}{A_e} & \delta_{12}^{LR}{m_{L}}{m_{E}} & \delta_{13}^{LR}{m_{L}}{m_{E}}  \\
   \delta_{12}^{LR}{m_{L}}{m_{E}} & m_{l_2}{A_e} & \delta_{23}^{LR}{m_{L}}{m_{E}}  \\
   \delta_{13}^{LR}{m_{L}}{m_{E}} & \delta_{23}^{LR}{m_{L}}{m_{E}} & m_{l_3}{A_e}  \\
\end{array}} \right){1\over {\upsilon_d}}.
\end{eqnarray}
Limited by the most relevant lepton flavor violating processes~\cite{sl-mix4}, we take the conservative choice for the off-diagonal parameters $\delta_{12}^{LL},\delta_{12}^{RR},\delta_{12}^{LR}\leq10^{-6}$, $\delta_{13}^{LL},\delta_{13}^{RR},\delta_{13}^{LR}\leq10^{-3}$ and $\delta_{23}^{LL},\delta_{23}^{RR},\delta_{23}^{LR}\leq10^{-2}$. So for simplicity, we will choose the off-diagonal parameters
\begin{eqnarray}
&&\delta_{23}^{LL}=\delta_{23}^{RR}=\delta_{23}^{LR}\equiv \delta_{23}^X, \nonumber\\ &&\delta_{13}^{LL}=\delta_{13}^{RR}=\delta_{13}^{LR}\equiv 10^{-1} \delta_{23}^X, \nonumber\\ &&\delta_{12}^{LL}=\delta_{12}^{RR}=\delta_{12}^{LR}\equiv 10^{-4} \delta_{23}^X.
\end{eqnarray}

Under above assumptions, ignoring the terms of the second order in $Y_{\nu}$ and assuming $(\upsilon_{\nu_i}^2+\upsilon_d^2-\upsilon_u^2)\approx (\upsilon_d^2-\upsilon_u^2)$, one can have the minimization conditions of the tree-level neutral scalar potential with respect to $\upsilon_{\nu_i}\:(i=1,2,3)$ below:
\begin{eqnarray}
m_{\tilde L_{ij}}^2 \upsilon_{\nu_j}+{G^2\over 4} (\upsilon_d^2-\upsilon_u^2)\upsilon_{\nu_i}=\Big[\lambda \upsilon_d (\upsilon_u^2+\upsilon_{\nu^c}^2) - \kappa \upsilon_u \upsilon_{\nu^c}^2\Big] Y_{\nu_i} -\upsilon_u \upsilon_{\nu^c}a_{\nu_i},
\label{eq-min}
\end{eqnarray}
where $G^2=g_1^2+g_2^2$ and $g_1 c_{_W} =g_2 s_{_W}=e$. Solving eq.~(\ref{eq-min}), we can gain the left-handed sneutrino VEVs
\begin{eqnarray}
\upsilon_{\nu_i}=\frac{{\rm{det}}\: T_i}{{\rm{det}}\: T},\qquad  (i=1,2,3),
\label{eq-vi}
\end{eqnarray}
where
\begin{eqnarray}
T = \left( {\begin{array}{*{20}{c}}
   m_{\tilde L_{11}}^2 +{G^2\over 4}(\upsilon_d^2-\upsilon_u^2) & m_{\tilde L_{12}}^2 & m_{\tilde L_{13}}^2  \\ [6pt]
   m_{\tilde L_{21}}^2 & m_{\tilde L_{22}}^2 +{G^2\over 4}(\upsilon_d^2-\upsilon_u^2) & m_{\tilde L_{23}}^2  \\ [6pt]
   m_{\tilde L_{31}}^2 & m_{\tilde L_{32}}^2 & m_{\tilde L_{33}}^2 +{G^2\over 4}(\upsilon_d^2-\upsilon_u^2)  \\ [6pt]
\end{array}} \right),
\end{eqnarray}
and $T_i$ can be acquired from $T$ by replacing the $i$-th column with
\begin{eqnarray}
\left( {\begin{array}{*{20}{c}}
   \Big[\lambda \upsilon_d (\upsilon_u^2+\upsilon_{\nu^c}^2) - \kappa \upsilon_u \upsilon_{\nu^c}^2\Big] Y_{\nu_1} -\upsilon_u \upsilon_{\nu^c}a_{\nu_1}  \\ [6pt]
   \Big[\lambda \upsilon_d (\upsilon_u^2+\upsilon_{\nu^c}^2) - \kappa \upsilon_u \upsilon_{\nu^c}^2\Big] Y_{\nu_2} -\upsilon_u \upsilon_{\nu^c}a_{\nu_2}  \\ [6pt]
   \Big[\lambda \upsilon_d (\upsilon_u^2+\upsilon_{\nu^c}^2) - \kappa \upsilon_u \upsilon_{\nu^c}^2\Big] Y_{\nu_3} -\upsilon_u \upsilon_{\nu^c}a_{\nu_3}  \\ [6pt]
\end{array}} \right).
\end{eqnarray}

Assuming that the charged lepton mass matrix in the flavor basis is in the diagonal form, we parameterize the unitary matrix which diagonalizes the effective light neutrino mass matrix $m_{eff}$ as~\cite{Uv1,Uv2,Uv3}
\begin{eqnarray}
{U_\nu} = \left( {\begin{array}{*{20}{c}}
   {{c_{12}}{c_{13}}} & {{s_{12}}{c_{13}}} & {s_{13}}  \\
   { - {s_{12}}{c_{23}} - {c_{12}}{s_{23}}{s_{13}}} & {{c_{12}}{c_{23}} - {s_{12}}}
   {s_{23}}{s_{13}} & {{s_{23}}{c_{13}}}  \\
   {s_{12}}{s_{23}} - {c_{12}}{c_{23}}{s_{13}} & { - {c_{12}}{s_{23}} - {s_{12}}}
   {c_{23}}{s_{13}} & {{c_{23}}{c_{13}}}  \\
\end{array}} \right),
\end{eqnarray}
where ${c_{ij}} = \cos {\theta _{ij}}$, ${s_{ij}} = \sin {\theta _{ij}}$, the angles
${\theta _{ij}} = \left[\: {0,\pi/2} \:\right]$, and CP violation phases are setting to zero, respectively. The unitary matrix $U_\nu$ diagonalizes $m_{eff}$ in the following way:
\begin{eqnarray}
U_\nu ^T m_{eff}^T{m_{eff}}{U_\nu} = diag({m_{\nu _1}^2},{m_{\nu _2}^2},{m_{\nu _3}^2}),
\label{neutrino-diagonalize}
\end{eqnarray}
In the case of 3-neutrino mixing, we have two possibilities on the neutrino mass spectrum~\cite{PDG}:
\begin{itemize}
\item (i) spectrum with normal ordering (NO):
\begin{eqnarray}
&&\;\:\,m_{\nu_1}<m_{\nu_2}<m_{\nu_3}, \nonumber\\
&&\Delta m_{\odot}^2 = m_{\nu_2}^2-m_{\nu_1}^2 >0,\quad\Delta m_{A}^2 = m_{\nu_3}^2-m_{\nu_1}^2>0;
\end{eqnarray}

\item (ii) spectrum with inverted ordering (IO):
\begin{eqnarray}
&&\;\:\,m_{\nu_3}<m_{\nu_1}<m_{\nu_2}, \nonumber\\
&&\Delta m_{\odot}^2 = m_{\nu_2}^2-m_{\nu_1}^2 >0, \quad\Delta m_{A}^2 = m_{\nu_3}^2-m_{\nu_2}^2 <0.
\end{eqnarray}
\end{itemize}

Limited on neutrino masses from neutrinoless double-$\beta$ decay~\cite{neu-m-limit} and cosmology~\cite{neu-m-limit1}, we choose the lightest neutrino mass $m_{\nu_{min}}=0.1\,{\rm{eV}}$ in the following. Through the experimental data on neutrino mass squared differences and mixing angles in eq.~(\ref{neu-oscillations1}) and  eq.~(\ref{neu-oscillations2}), we can obtain the values of $\theta_{ij}$ and other two neutrino masses. The effective light neutrino mass matrix $m_{eff}$ can approximate as~\cite{meu-m}
\begin{eqnarray}
{m_{ef{f_{ij}}}} \approx \frac{{2A{\upsilon_{\nu^c}}}}{{3\Delta }}{b_i}{b_j} + \frac{{1 - 3{\delta _{ij}}}}{{6\kappa {\upsilon_{\nu^c}}}}{a_i}{a_j},
\end{eqnarray}
where
\begin{eqnarray}
&&\Delta  = {\lambda ^2}{(\upsilon_d^2 + \upsilon_u^2)}^2 + 4\lambda \kappa {\upsilon_{\nu^c}^2}{\upsilon_d}{\upsilon_u} - 12{\lambda ^2}{\upsilon_{\nu^c}}AB,\nonumber\\
&& A = \kappa {\upsilon_{\nu^c}^2} + \lambda {\upsilon_d}{\upsilon_u},\quad \frac{1}{B}= \frac{e^2}{c_{_W}^2{M_1}} + \frac{e^2}{s_{_W}^2{M_2}} , \nonumber\\
&&{a_i} = {Y_{{\nu _i}}}{\upsilon_u}\:, \qquad\qquad\;\: {b_i} = {Y_{{\nu _i}}}{\upsilon_d} + 3\lambda \upsilon_{\nu_i}.
\end{eqnarray}
Then, we can numerically derive $Y_{\nu_i} \sim \mathcal{O}(10^{-7})$ and $a_{\nu_i} \sim \mathcal{O}(-10^{-4}{\rm{GeV}})$ from eq.~(\ref{neutrino-diagonalize}). Accordingly, $\upsilon_{\nu_i} \sim \mathcal{O}(10^{-4}{\rm{GeV}})$ through eq.~(\ref{eq-vi}). So $\upsilon_{\nu_i}\ll\upsilon_{u,d}$, then we can have
\begin{eqnarray}
\tan\beta\simeq \frac{\upsilon_u}{\upsilon_d}.
\end{eqnarray}

Through the above analysis, assuming universality for the soft breaking bino and wino masses, $M_1= 0.5 \,M_2$, the free parameters affect our next analysis are
\begin{eqnarray}
\tan \beta , \:  \kappa ,\: \lambda , \: {\upsilon_{\nu^c}},\:{A_\lambda },  \:{A_e}, \:{m_L},\:{m_E},\:M_2,\:\delta_{23}^X.
\end{eqnarray}

\subsection{Transition magnetic moment of Majorana neutrinos}

\begin{figure}
\setlength{\unitlength}{1mm}
\centering
\begin{minipage}[c]{0.5\textwidth}
\includegraphics[width=2.9in]{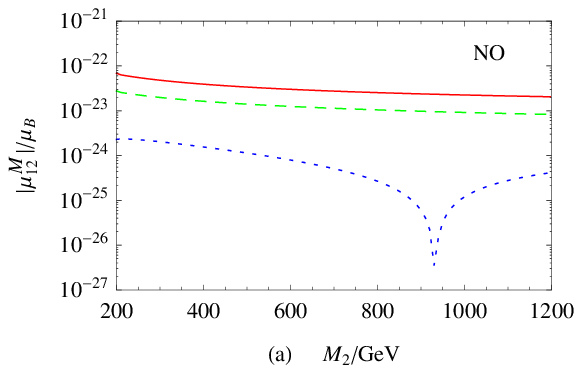}
\end{minipage}%
\begin{minipage}[c]{0.5\textwidth}
\includegraphics[width=2.9in]{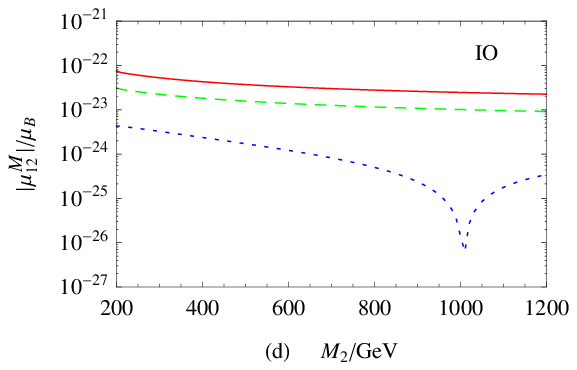}
\end{minipage}
\begin{minipage}[c]{0.5\textwidth}
\includegraphics[width=2.9in]{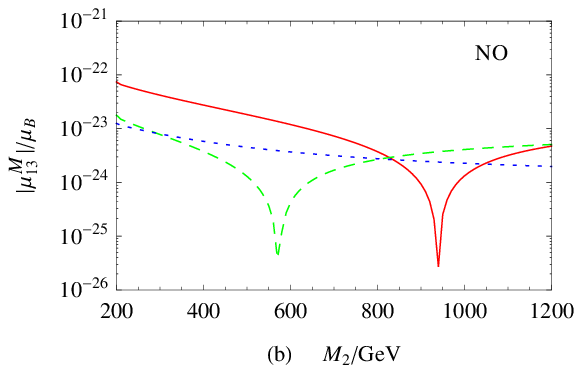}
\end{minipage}%
\begin{minipage}[c]{0.5\textwidth}
\includegraphics[width=2.9in]{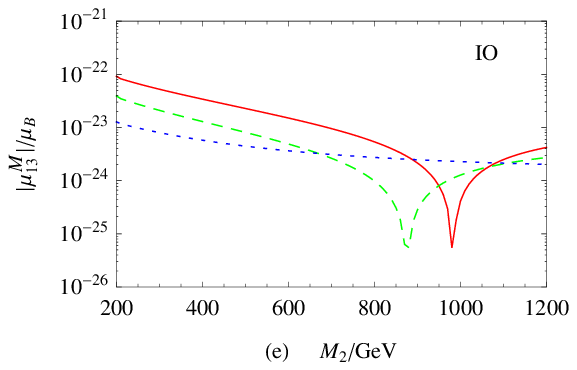}
\end{minipage}
\begin{minipage}[c]{0.5\textwidth}
\includegraphics[width=2.9in]{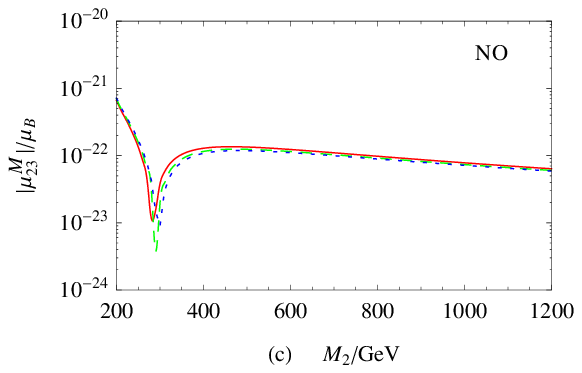}
\end{minipage}%
\begin{minipage}[c]{0.5\textwidth}
\includegraphics[width=2.9in]{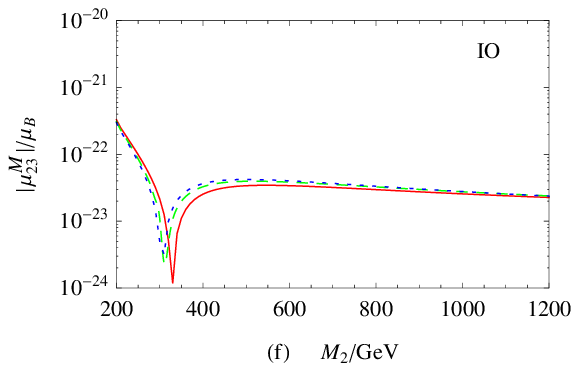}
\end{minipage}
\caption[]{(Color online) Assuming neutrino mass spectrum with NO (a, b, c) or IO (d, e, f) in the $\mu\nu$SSM, we plot the transition magnetic moment of Majorana neutrinos varying with $M_2$, as $\delta_{23}^X=10^{-2}$ (solid line), $\delta_{23}^X=0.5\times10^{-2}$ (dashed line) and $\delta_{23}^X=0$ (dotted line), respectively.}
\label{fig-u}
\end{figure}

Recently, a neutral Higgs with mass $m_{h}\sim 124-126\;{\rm GeV}$ was reported by ATLAS~\cite{ATLAS} and CMS~\cite{CMS}, which contributes a strict constraint on relevant parameter space of the model. In the $\mu\nu{\rm SSM}$, one can obtain a tree-level upper bound on the SM-like Higgs mass~\cite{mnSSM1}:
\begin{eqnarray}
m_h^2\leq m_Z^2 \cos^2 2\beta + {{6 \lambda^2 s_{_W}^2 c_{_W}^2}\over e^2} m_Z^2 \sin^2 2\beta,
\label{mh-upper}
\end{eqnarray}
where $m_Z$ is the mass of $Z$-boson. Compared with the MSSM, the second term of the right side in eq.~(\ref{mh-upper}) is additional contribution to the SM-like Higgs mass. So, the SM-like Higgs in the $\mu\nu{\rm SSM}$ can easily account for the mass around $125\,{\rm GeV}$, through the main radiative corrections from the top quark and its supersymmetric partners. Also due to this, the $\mu\nu{\rm SSM}$ favors small $\tan \beta$~\cite{mnSSM2,Zhanghb}. Hence, we will take $\tan \beta=3$ for simplicity in the following.

And we choose the relevant parameters as default in our numerical calculation for convenience:
\begin{eqnarray}
&&\kappa=0.01, \quad {A_\lambda }=\upsilon_{\nu^c}=500\:{\rm{GeV}},   \nonumber\\
&&\lambda=0.1, \quad\;\;  A_e= m_L= m_E=200\:{\rm{GeV}}.
\end{eqnarray}
Here we take relatively small values of the parameters ${\kappa}$ and $\lambda$, imposing the Landau pole condition at the high-energy scale~\cite{mnSSM1,mnSSM2}. Limited on supersymmetric particle masses from the Particle Data Group~\cite{PDG}, we choose $A_e=m_{L,E}=200\:{\rm{GeV}}$ and $M_2\geq200\:{\rm{GeV}}$ in the next numerical calculation.

With the above assumptions on parameter space of the $\mu\nu$SSM, in figure~\ref{fig-u} we plot the transition magnetic moment of Majorana neutrinos varying with $M_2$ assuming neutrino mass spectrum with NO or IO, as $\delta_{23}^X=10^{-2}$ (solid line), $\delta_{23}^X=0.5\times10^{-2}$ (dashed line) and $\delta_{23}^X=0$ (dotted line), respectively. One can find that the general trend on the Majorana neutrino transition magnetic moment is decreasing, along with increasing of $M_2$. Here, the wino-like chargino mass is dependent on soft breaking wino mass $M_2$. When $M_2$ is large, the wino-like chargino mass will be large, which causes that the wino-like chargino loop contribution is suppressed. Conversely, if $M_2$ is enough small, we could obtain relatively large transition magnetic moment.

However, in figure~\ref{fig-u} we also can see that transition magnetic moment of Majorana neutrinos may have a sharp decrease in some parameter space. Due to the fact that Majorana neutrino coincides with its antiparticle, the Majorana neutrino transition magnetic moment $\mu_{ij}^M$ in eq.~(\ref{Majorana-MDM}) contains the Dirac-neutrino-like term $\mu_{ij}^D$ and the Dirac-antineutrino-like term $-\mu_{ji}^D$. Considered the supersymmetric particle loop contributions in the $\mu\nu{\rm SSM}$, the Majorana neutrino transition magnetic moment could have a resonating absorption in some parameter space, which originates from the interference between the Dirac-neutrino-like term and the Dirac-antineutrino-like term.

In this paragraph, we will analyse the contribution from the off-diagonal parameter $\delta_{23}^X$. In figure~\ref{fig-u}, the transition magnetic moment $\mu_{23}^M$ doesn't change much for different values of the off-diagonal parameter $\delta_{23}^X$. However, the transition magnetic moments $\mu_{12}^M$ and $\mu_{13}^M$ could be enhanced largely, with increasing of $\delta_{23}^X$. It's well known that the transition magnetic moment is flavor dependent. Taking into account the off-diagonal terms of soft breaking slepton mass matrices and trilinear coupling matrices enlarges slepton flavor mixing, which could give a contribution to the transition magnetic moment through the charged fermions and charged scalars loop diagrams.

Compared normal ordering with inverted ordering for neutrino mass spectrum, the Majorana neutrino transition magnetic moments depicted in figure~\ref{fig-u} are somewhat alike, where the lightest neutrino mass is the same. But, we also can find some differences between normal ordering and inverted ordering. First, the value of $M_2$ for a resonating absorption is different. And the transition magnetic moment $\mu_{23}^M$ for normal ordering in figure~\ref{fig-u}(c) is relatively larger than that for inverted ordering in figure~\ref{fig-u}(f).

In the Standard Model augmented by nonzero Majorana neutrino masses, the Majorana neutrino transition magnetic moments $\mu_{ij}^M\sim\mathcal{O}(10^{-24}\mu_B)$, when the lightest neutrino mass $m_{\nu_{min}}=0.1\,{\rm{eV}}$~\cite{Broggini}. In figure~\ref{fig-u}, the Majorana neutrino transition magnetic moments in the $\mu\nu$SSM could be enhanced to $\mathcal{O}(10^{-21,-22}\mu_B)$, for small $M_2$ and large $\delta_{23}^X$. Here the supersymmetric particle loops give the dominant contributions to the transition magnetic moments. In the oscillation of supernova neutrinos, Majorana neutrino transition magnetic moments
reveal that moments as small as $10^{-24}\mu_B$ may leave a potentially observable imprint on the energy spectra of neutrinos and antineutrinos from supernovae~\cite{neu-exp1,neu-exp2,neu-exp3}. So, the relatively large transition magnetic moment of Majorana neutrinos in the $\mu\nu$SSM will be more easily detected than that in the Standard Model augmented by massive Majorana neutrinos.

\section{Conclusions}
\label{sec5}

The $\mu\nu$SSM, one supersymmetric extension of the Standard Model, can generate three tiny massive Majorana neutrinos through the seesaw mechanism. Applying effective Lagrangian method and on-shell scheme, we investigate the transition magnetic moment of Majorana neutrinos in the model. Under the present neutrino experimental constraints and some assumptions of the parameter space, we consider the two possibilities on the neutrino mass spectrum with normal or inverted ordering.

The numerical results show that the transition magnetic moment of Majorana neutrinos in the $\mu\nu$SSM can be enhanced to $\mathcal{O}(10^{-21,-22}\mu_B)$, when the supersymmetric particles are light and the off-diagonal terms of soft breaking slepton parameters are large. In the oscillation of supernova neutrinos, Majorana neutrino transition magnetic moments reveal that moments as small as $10^{-24}\mu_B$ may leave a potentially observable from supernovae~\cite{neu-exp1,neu-exp2,neu-exp3}. Therefore, the Majorana neutrino transition magnetic moments in the $\mu\nu$SSM have more opportunities to be detected from supernovae in the future.

\section*{Acknowledgements}

The work has been supported by the National Natural Science Foundation of China (NNSFC)
with Grant No. 11275036, No. 11047002, the open project of State
Key Laboratory of Mathematics-Mechanization with Grant No. Y3KF311CJ1, the Natural
Science Foundation of Hebei province with Grant No. A2013201277, and Natural Science Fund of Hebei University with Grant No. 2011JQ05, No. 2012-242.

\appendix

\section{Mass Matrices\label{appendix-mass}}
In this appendix, we give the relative mass matrices for this study in the $\mu\nu$SSM.

\subsection{Charged scalar mass matrix}
The quadratic potential includes
\begin{eqnarray}
{V_{quadratic}} = {S'^{-T}}M_{S^\pm}^2S'^+ ,
\end{eqnarray}
where ${S'^{ \pm T}} = (H_d^ \pm ,H_u^ \pm ,\tilde e_{L_i}^ \pm ,\tilde e_{R_i}^ \pm )$ is in the unrotated basis, $\tilde e_{L_i}^- \equiv \tilde e_i$ and $\tilde e_{R_i}^+ \equiv \tilde e_i^c$. The expressions for the independent coefficients of $M_{S^\pm}^2$ are given in detail below:
\begin{eqnarray}
&&M_{H_d^\pm H_d^\pm}^2 = m_{H_d}^2 + \frac{g_2^2}{2}(\upsilon_u^2-\upsilon_{\nu_i}\upsilon_{\nu_i}) + \frac{G^2}{4}(\upsilon_d^2-\upsilon_u^2+\upsilon_{\nu_i}\upsilon_{\nu_i})+\lambda_i \lambda_j \upsilon_{\nu_i^c}\upsilon_{\nu_j^c} \quad \nonumber\\
&&\qquad\qquad\;\;\: + \: Y_{e_{ik}}Y_{e_{jk}}\upsilon_{\nu_i}\upsilon_{\nu_j} , \\
&&M_{H_u^\pm H_u^\pm}^2 = m_{H_u}^2 + \frac{g_2^2}{2}(\upsilon_d^2+\upsilon_{\nu_i}\upsilon_{\nu_i}) - \frac{G^2}{4}(\upsilon_d^2 - \upsilon_u^2 + \upsilon_{\nu_i}\upsilon_{\nu_i})+\lambda_i \lambda_j \upsilon_{\nu_i^c}\upsilon_{\nu_j^c} \nonumber\\
&&\qquad\qquad\;\;\: + \: Y_{\nu_{ik}}Y_{\nu_{ij}}\upsilon_{\nu_j^c}\upsilon_{\nu_k^c} , \\
&&M_{H_d^\pm H_u^\pm}^2 = (A_\lambda \lambda)_i \upsilon_{\nu_i^c} + \frac{g_2^2}{2}\upsilon_d \upsilon_u - \lambda_i \lambda_i \upsilon_d \upsilon_u  +\lambda_k \kappa_{ijk}\upsilon_{\nu_i^c}\upsilon_{\nu_j^c} +  Y_{\nu_{ij}} \lambda_j \upsilon_u \upsilon_{\nu_i} , \\
&&M_{H_d^\pm \tilde e_{L_i}^\pm}^2 = \frac{g_2^2}{2}\upsilon_d \upsilon_{\nu_i} - Y_{\nu_{ij}}\lambda_k \upsilon_{\nu_k^c} \upsilon_{\nu_j^c} -  Y_{e_{ij}}Y_{e_{kj}}\upsilon_d \upsilon_{\nu_k},\\
&&M_{H_u^\pm \tilde e_{L_i}^\pm}^2 = \frac{g_2^2}{2}\upsilon_u \upsilon_{\nu_i} -  {(A_\nu Y_\nu)}_{ij} \upsilon_{\nu_j^c} + Y_{\nu_{ij}}\lambda_j \upsilon_d \upsilon_u - Y_{\nu_{ij}} \kappa_{ljk} \upsilon_{\nu_l^c}\upsilon_{\nu_k^c} \nonumber\\
&&\qquad\qquad\;\;\: -\:  Y_{\nu_{ik}} Y_{\nu_{kj}}  \upsilon_u \upsilon_{\nu_j} , \\
&&M_{H_d^\pm \tilde e_{R_i}^\pm}^2 = -(A_e Y_e)_{ji} \upsilon_{\nu_j}  -  Y_{e_{ki}}Y_{\nu_{kj}}\upsilon_u \upsilon_{\nu_j^c} , \\
&&M_{H_u^\pm \tilde e_{R_i}^\pm}^2 = - Y_{e_{ki}} (\lambda_j \upsilon_{\nu_j^c}\upsilon_{\nu_k} +Y_{\nu_{kj}}  \upsilon_d \upsilon_{\nu_j^c}) , \\
&&M_{\tilde e_{L_i}^\pm \tilde e_{L_j}^\pm}^2 = m_{\tilde L_{ij}}^2 + \frac{1}{4}(g_1^2-g_2^2)(\upsilon_d^2 - \upsilon_u^2 + \upsilon_{\nu_k}\upsilon_{\nu_k})\delta_{ij} + \frac{g_2^2}{2}\upsilon_{\nu_i} \upsilon_{\nu_j}  \nonumber\\
&&\qquad\qquad\;\;\: +\: Y_{\nu_{il}} Y_{\nu_{jk}}  \upsilon_{\nu_l^c}\upsilon_{\nu_k^c} +  Y_{e_{ik}} Y_{e_{jk}}  \upsilon_d^2  , \\
&&M_{\tilde e_{L_i}^\pm \tilde e_{R_j}^\pm}^2 = {(A_e Y_e)}_{ij}\upsilon_d - Y_{e_{ij}} \lambda_k  \upsilon_u \upsilon_{\nu_k^c}   , \\
&&M_{\tilde e_{R_i}^\pm \tilde e_{R_j}^\pm}^2 =  m_{\tilde e_{ij}^c}^2 - \frac{1}{2}g_1^2(\upsilon_d^2 - \upsilon_u^2 + \upsilon_{\nu_k}\upsilon_{\nu_k})\delta_{ij}  +  Y_{e_{ki}}Y_{e_{kj}} \upsilon_d^2  + Y_{e_{li}}Y_{e_{kj}}\upsilon_{\nu_k}\upsilon_{\nu_l} .
\end{eqnarray}
Here, imposing the minimization conditions of the tree-level neutral scalar potential with respect to $\upsilon_d$ and $\upsilon_u$, one can have
\begin{eqnarray}
&&m_{{H_d}}^2=\Big\{ -\frac{G^2}{4}( \upsilon_d^2 - \upsilon_u^2 +\upsilon_{\nu_i}\upsilon_{\nu_i}) \upsilon_d  + (A_\lambda \lambda)_i {\upsilon_u} \upsilon_{\nu_i^c} + {\lambda _j}{\kappa _{ijk}}{\upsilon_u}\upsilon_{\nu_i^c} \upsilon_{\nu_k^c}  \nonumber\\
&&\qquad\quad\quad\;  - \:({\lambda _i}{\lambda _j}\upsilon_{\nu_i^c}\upsilon_{\nu_j^c}  + {\lambda _i}{\lambda _i}\upsilon_u^2){\upsilon_d}  + {Y_{{\nu_{ij}}}}\upsilon_{\nu_i}({\lambda _k}\upsilon_{\nu_k^c}\upsilon_{\nu_j^c} + {\lambda _j}\upsilon_u^2)\Big\}{1\over \upsilon_d},\\
&&m_{{H_u}}^2=\Big\{ \frac{{G^2}}{4}(\upsilon_d^2 - \upsilon_u^2 +\upsilon_{\nu_i}\upsilon_{\nu_i})\upsilon_u  - {(A_\nu Y_\nu)}_{ij}\upsilon_{\nu_i}\upsilon_{\nu_j^c} + (A_\lambda \lambda)_i {\upsilon_d} \upsilon_{\nu_i^c}  \nonumber\\
&&\qquad\quad\quad\;  - \:({\lambda _i}{\lambda _j}\upsilon_{\nu_i^c}\upsilon_{\nu_j^c}  - {\lambda _i}{\lambda _i}\upsilon_u^2){\upsilon_u} - {Y_{{\nu_{ij}}}}\upsilon_{\nu_i}({\kappa _{ljk}}\upsilon_{\nu_l^c} \upsilon_{\nu_k^c} - 2 {\lambda _j}\upsilon_d \upsilon_u )  \nonumber\\
&&\qquad\quad\quad\; + \: {\lambda _j}{\kappa _{ijk}}{\upsilon_d}\upsilon_{\nu_i^c} \upsilon_{\nu_k^c} - ( Y_{\nu_{ki}}Y_{\nu_{kj}}\upsilon_{\nu_i^c}\upsilon_{\nu_j^c}  + Y_{\nu_{ik}}Y_{\nu_{jk}}\upsilon_{\nu_i}\upsilon_{\nu_j} )  \upsilon_u \Big\} {1\over{\upsilon_u}}.
\end{eqnarray}

We can use an $8\times8$ unitary matrix $R_{S^\pm}$ to diagonalize the mass matrix $M_{S^\pm}^2$
\begin{eqnarray}
R_{S^\pm}^TM_{S^\pm}^2{R_{S^\pm}} = {(M_{S^\pm}^{diag})^2}.
\end{eqnarray}
Then, $S'^ \pm _\alpha$ can be rotated to the mass eigenvectors $S^ \pm _\alpha $ (${\alpha  = 1, \ldots, 8}$):
\begin{eqnarray}
H_d^ \pm  = R_{{S^ \pm }}^{1\alpha }S_\alpha ^ \pm ,\quad H_u^ \pm = R_{{S^ \pm }}^{2\alpha }S_\alpha ^ \pm ,\quad \tilde e_{L_i}^ \pm =R_{{S^ \pm }}^{(2 + i)\alpha }S_\alpha ^ \pm ,\quad \tilde e_{R_i}^ \pm = R_{{S^ \pm }}^{(5 + i)\alpha }S_\alpha ^ \pm.
\end{eqnarray}

\subsection{Neutral fermion mass matrix}
Neutrinos mix with the neutralinos and in the basis ${\chi '^{ \circ T}} = \left( {{{\tilde B}^ \circ },{{\tilde W}^ \circ },{{\tilde H}_d}{\rm{,}}{{\tilde H}_u},{\nu_{R_i}},{\nu_{L{_i}}}} \right)$, we can obtain the neutral fermion mass terms in the Lagrangian:
\begin{eqnarray}
\mathcal{L}_{mass} = - \frac{1}{2}{\chi '^{ \circ T}}{M_n}{\chi '^ \circ } + {\rm{H.c.}}\:  ,
\end{eqnarray}
where
\begin{eqnarray}
{M_n} = \left( {\begin{array}{*{20}{c}}
   M & {{m^T}}  \\
   m & {{0_{3 \times 3}}}  \\
\end{array}} \right),
\end{eqnarray}
with
\begin{eqnarray}
m = \left( {\begin{array}{*{20}{c}}
   {  -\frac{g_1}{\sqrt 2 }\upsilon_{{\nu _1}}} & { \frac{g_2}{\sqrt 2 }\upsilon_{{\nu _1}}} & 0 & {{Y_{{\nu _{1i}}}}{\upsilon_{\nu _i^c}}} & {{Y_{{\nu _{11}}}}{\upsilon_u}} & {{Y_{{\nu _{12}}}}{\upsilon_u}} & {{Y_{{\nu _{13}}}}{\upsilon_u}}  \\
   {  -\frac{g_1}{\sqrt 2 }\upsilon_{{\nu _2}}} & { \frac{g_2}{\sqrt 2 }\upsilon_{{\nu _2}}} & 0 & {{Y_{{\nu _{2i}}}}{\upsilon_{\nu _i^c}}} & {{Y_{{\nu _{21}}}}{\upsilon_u}} & {{Y_{{\nu _{22}}}}{\upsilon_u}} & {{Y_{{\nu _{23}}}}{\upsilon_u}}  \\
   {  -\frac{g_1}{\sqrt 2 }\upsilon_{{\nu _3}}} & { \frac{g_2}{\sqrt 2 }\upsilon_{{\nu _3}}} & 0 & {{Y_{{\nu _{3i}}}}{\upsilon_{\nu _i^c}}} & {{Y_{{\nu _{31}}}}{\upsilon_u}} & {{Y_{{\nu _{32}}}}{\upsilon_u}} & {{Y_{{\nu _{33}}}}{\upsilon_u}}  \\
\end{array}} \right)
\end{eqnarray}
and
\begin{eqnarray}
&&M = \left( {\begin{array}{*{20}{c}}
   {{M_1}} & 0 & {\frac{-g_1}{{\sqrt 2 }}{\upsilon _d}} & {\frac{g_1}{{\sqrt 2 }}{\upsilon _u}} & 0 & 0 & 0  \\
   0 & {{M_2}} & {\frac{g_2}{{\sqrt 2 }}{\upsilon _d}} & {\frac{-g_2}{{\sqrt 2 }}{\upsilon _u}} & 0 & 0 & 0  \\
   {\frac{-g_1}{{\sqrt 2 }}{\upsilon _d}} & {\frac{g_2}{{\sqrt 2 }}{\upsilon _d}} & 0 & {-{\lambda _i}{\upsilon _{\nu _i^c}}} & { - {\lambda _1}{\upsilon _u}} & { - {\lambda _2}{\upsilon _u}} & { - {\lambda _3}{\upsilon _u}}  \\
   {\frac{g_1}{{\sqrt 2 }}{\upsilon _u}} & {\frac{-g_2}{{\sqrt 2 }}{\upsilon _u}} & {-{\lambda _i}{\upsilon _{\nu _i^c}}} & 0 & {y_1} & {y_2} & { y_3}  \\
   0 & 0 & { - {\lambda _1}{\upsilon _u}} & { y_1} & {2{\kappa _{11j}}{\upsilon _{\nu _j^c}}} & {2{\kappa _{12j}}{\upsilon _{\nu _j^c}}} & {2{\kappa _{13j}}{\upsilon _{\nu _j^c}}}  \\
   0 & 0 & { - {\lambda _2}{\upsilon _u}} & { y_2} & {2{\kappa _{21j}}{\upsilon _{\nu _j^c}}} & {2{\kappa _{22j}}{\upsilon _{\nu _j^c}}} & {2{\kappa _{23j}}{\upsilon _{\nu _j^c}}}  \\
   0 & 0 & { - {\lambda _3}{\upsilon _u}} & { y_3} & {2{\kappa _{31j}}{\upsilon _{\nu _j^c}}} & {2{\kappa _{32j}}{\upsilon _{\nu _j^c}}} & {2{\kappa _{33j}}{\upsilon _{\nu _j^c}}}  \\
\end{array}} \right) ,
\end{eqnarray}
where $y_i=- {\lambda _i}{\upsilon _d}+ {{Y_{{\nu _{ji}}}}{\upsilon _{{\nu _j}}} }$. Here, the submatrix $m$ is neutralino-neutrino mixing, and the submatrix $M$ is neutralino mass matrix. This $10\times10$ symmetric matrix $M_n$ can be diagonalized by a $10\times10$ unitary matrix $Z_n$:
\begin{eqnarray}
Z_n^T{M_n}{Z_n} = {M_{nd}},
\end{eqnarray}
where $M_{nd}$ is the diagonal neutral fermion mass matrix. Then, we have the neutral fermion mass eigenstates:
\begin{eqnarray}
\chi _\alpha ^ \circ  = \left( {\begin{array}{*{20}{c}}
   {\kappa _\alpha ^ \circ }  \\
   { \overline{\kappa _\alpha ^ \circ} }  \\
\end{array}} \right), \quad {\alpha  = 1, \ldots, 10}
\end{eqnarray}
with
\begin{eqnarray}
\left\{ {\begin{array}{*{20}{c}}
   {{\tilde B^ \circ } = Z_n^{1\alpha }\kappa _\alpha ^ \circ ,\quad\: {\tilde H_d} = Z_n^{3\alpha }\kappa _\alpha ^ \circ ,\quad{\nu_{R_i}} = Z_n^{\left( {4 + i} \right)\alpha }\kappa _\alpha ^ \circ ,\,}  \\
   {{\tilde W^ \circ } = Z_n^{2\alpha }\kappa _\alpha ^ \circ ,\quad{\tilde H_u} = Z_n^{4\alpha }\kappa _\alpha ^ \circ , \quad{\nu_{L_i}} = Z_n^{\left( {7 + i} \right)\alpha }\kappa _\alpha ^ \circ .\:}  \\
\end{array}} \right.
\end{eqnarray}

\subsection{Charged fermion mass matrix}
Charged leptons mix with the charginos and therefore in the unrotated basis where ${\Psi ^{ - T}} = \left( { - i{{\tilde \lambda }^ - },\tilde H_d^ - ,e_{L{_i}}^ - } \right)$ and ${\Psi ^{ + T}} = \left( { - i{{\tilde \lambda }^ + },\tilde H_u^ + ,e_{R{_i}}^+} \right)$, one can have the charged fermion mass terms in the Lagrangian:
\begin{eqnarray}
\mathcal{L}_{mass} = - {\Psi ^{ - T}}{M_c}{\Psi^+} + {\rm{H.c.}}\:,
\end{eqnarray}
where
\begin{eqnarray}
{M_c} = \left( {\begin{array}{*{20}{c}}
   {{M_ \pm }} & b  \\
   c & {{m_l}}  \\
\end{array}} \right).
\end{eqnarray}
Here, the submatrix $M_ \pm $ is chargino mass matrix
\begin{eqnarray}
{M_ \pm } = \left( {\begin{array}{*{20}{c}}
   {{M_2}} & {g_2 {\upsilon_u}}  \\
   {g_2 {\upsilon_d}} & {{\lambda _i} \upsilon_{\nu_i^c}}  \\
\end{array}} \right).
\end{eqnarray}
And the submatrices $b$ and $c$ give rise to chargino-charged lepton mixing. They are defined as
\begin{eqnarray}
b = \left( {\begin{array}{*{20}{c}}
   0 & 0 & 0  \\
   { - {Y_{e_{i1}}} \upsilon_{\nu _i}} & { - {Y_{e_{i2}}} \upsilon_{\nu _i}} & { - {Y_{e_{i3}}} \upsilon_{\nu _i}}  \\
\end{array}} \right),
\end{eqnarray}
\begin{eqnarray}
c = \left( {\begin{array}{*{20}{c}}
   {g_2 \upsilon_{\nu _1}} & { - {Y_{\nu_{1i}}}\upsilon_{\nu_i^c}}  \\
   {g_2 \upsilon_{\nu _2}} & { - {Y_{\nu_{2i}}}\upsilon_{\nu_i^c}}  \\
   {g_2 \upsilon_{\nu _3}} & { - {Y_{\nu_{3i}}}\upsilon_{\nu_i^c}}  \\
\end{array}} \right).
\end{eqnarray}
And the submatrix $m_l$ is the charged lepton mass matrix
\begin{eqnarray}
{m_l} = \left( {\begin{array}{*{20}{c}}
   {{Y_{e_{11}}}{\upsilon_d}} & {{Y_{e_{12}}}{\upsilon_d}} & {{Y_{e_{13}}}{\upsilon_d}}  \\
   {{Y_{e_{21}}}{\upsilon_d}} & {{Y_{e_{22}}}{\upsilon_d}} & {{Y_{e_{23}}}{\upsilon_d}}  \\
   {{Y_{e_{31}}}{\upsilon_d}} & {{Y_{e_{32}}}{\upsilon_d}} & {{Y_{e_{33}}}{\upsilon_d}}  \\
\end{array}} \right).
\end{eqnarray}
This $5\times5$ mass matrix $M_c$ can be diagonalized by the $5\times5$ unitary matrices $Z_-$ and $Z_+$:
\begin{eqnarray}
Z_ - ^T{M_c}{Z_ + } = {M_{cd}},
\end{eqnarray}
where $M_{cd}$ is the diagonal charged fermion mass matrix. Then, one can obtain the charged fermion mass eigenstates:
\begin{eqnarray}
{\chi _\alpha } = \left( {\begin{array}{*{20}{c}}
   {\kappa _\alpha ^ - }  \\
   {\overline{{\kappa _\alpha ^+}}}  \\
\end{array}} \right),\quad {\alpha  = 1, \ldots, 5}
\end{eqnarray}
with
\begin{eqnarray}
\left\{ {\begin{array}{*{20}{c}}
   {{{\tilde \lambda }^ - } = iZ_ - ^{1\alpha }\kappa _\alpha ^ - ,\quad \tilde H_d^ -  = Z_ - ^{2\alpha }\kappa _\alpha ^ - ,\quad {e_{L_i}} = Z_ - ^{\left( {2 + i} \right)\alpha }\kappa _\alpha ^ - ;}  \\
   \;{{{\tilde \lambda }^ + } = iZ_ + ^{1\alpha }\kappa _\alpha ^ + ,\quad {{\tilde H}_u^+} = Z_ + ^{2\alpha }\kappa _\alpha ^ + ,\quad {e_{R_i}} = Z_+^{\left( {2 + i} \right)\alpha }\kappa _\alpha ^+.\:}  \\
\end{array}} \right.
\end{eqnarray}

\section{Couplings\label{appendix-coupling}}

In this appendix, we show the relevant couplings in the computation of the neutrino magnetic moment within framework of the $\mu\nu$SSM. And we use the indices $i,j=1,2,3$, $\beta =1,\cdots,5$, $\alpha =1,\cdots,8$ and $\eta=1,\cdots,10$.

\subsection{Charged fermion-neutral fermion-gauge boson}
The couplings of Charged fermion, neutral fermion and gauge boson are given by
\begin{eqnarray}
\mathcal{L} = &&e F_\mu \bar{\chi}_\beta \gamma^\mu \chi_\beta  + W_\mu^+ \bar{\chi}_\eta^\circ (C_L^{W \chi_\beta \bar{\chi}_\eta^\circ}\gamma^\mu P_L +  C_R^{W \chi_\beta \bar{\chi}_\eta^\circ}\gamma^\mu P_R) \chi_\beta  \nonumber\\
&& + W_\mu^- \bar{\chi}_\beta (C_L^{W \chi_\eta^\circ \bar{\chi}_\beta}\gamma^\mu P_L +  C_R^{W \chi_\eta^\circ \bar{\chi}_\beta}\gamma^\mu P_R) \chi_\eta^\circ + \cdots,
\end{eqnarray}
where the coefficients are
\begin{eqnarray}
&&C_L^{W{\chi _{^\beta }}\bar \chi _\eta ^ \circ } =  - \frac{e}{{\sqrt 2 {s_{_W}}}} \Big( {\sqrt 2 Z_ - ^{1\beta }Z{{_n^{2\eta }}^ * } + Z_ - ^{2\beta }Z{{_n^{3\eta }}^ * } + Z_ - ^{(2 + i)\beta }Z{{_n^{(7 + i)\eta }}^ * }} \Big),\\
&&C_R^{W{\chi _{^\beta }}\bar \chi _\eta ^ \circ } =  - \frac{e}{{\sqrt 2 {s_{_W}}}} \Big( \sqrt 2 Z{{_ + ^{1\beta }}^ * }Z_n^{2\eta } - Z{{_ + ^{2\beta }}^ * }Z_n^{4\eta } \Big),\\
&&C_L^{W\chi _\eta ^ \circ {{\bar \chi }_{^\beta }}} = \Big( C_L^{W{\chi _{^\beta }}\bar \chi _\eta ^ \circ }\Big)^*,\qquad\quad   C_R^{W\chi _\eta ^ \circ {{\bar \chi }_{^\beta }}} = \Big( C_R^{W{\chi _{^\beta }}\bar \chi _\eta ^ \circ }\Big)^*.
\end{eqnarray}

\subsection{Charged fermion-neutral fermion-charged scalar}
The couplings of charged fermion, neutral fermion and charged scalar are similarly written as
\begin{eqnarray}
\mathcal{L} = &&S_\alpha^- \bar{\chi}_\beta (C_L^{S_\alpha ^ - \chi _\eta ^ \circ {{\bar \chi }_\beta }}{P_L} + C_R^{S_\alpha ^ - \chi _\eta ^ \circ {{\bar \chi }_\beta }}{P_R} ) \chi_\eta^\circ \nonumber\\
&&+ \: S_\alpha^{-\ast} \bar{\chi}_\eta^\circ (C_L^{S_\alpha ^{-\ast} {\chi _\beta }\bar \chi _\eta ^ \circ }{P_L} + C_R^{S_\alpha ^{-\ast} {\chi _\beta }\bar \chi _\eta ^ \circ }{P_R} ) \chi_\beta  + \cdots.
\end{eqnarray}
And the coefficients are
\begin{eqnarray}
&&C_L^{S_\alpha^- \chi _\eta^\circ {{\bar{\chi}}_\beta }} =   \frac{-e}{{\sqrt{2} {s_{_W}}{c_{_W}}}}R{_{{S^\pm }}^{2\alpha \ast } }Z_+^{2\beta} \Big( {{c_{_W}}Z_n^{2\eta } + {s_W}Z_n^{1\eta }} \Big) - \frac{{\sqrt{2} e}}{{{s_{_W}}}}R{_{{S^\pm }}^{(5 + i)\alpha\ast } }Z_ + ^{(2 + i)\beta }Z_n^{1\eta }\nonumber\\
&&\qquad\qquad\quad\;\; - \frac{e}{{{s_{_W}}}}R{_{{S^ \pm }}^{2\alpha\ast } }Z_ + ^{1\beta }Z_n^{4\eta }  + {Y_{e_{ij}}}Z_ + ^{(2 + j)\beta } \Big( R_{{S^ \pm }}^{1\alpha }Z_n^{(7 + i)\eta } - R_{{S^ \pm }}^{(2 + i)\alpha }Z_n^{3\eta } \Big) \nonumber\\
&&\qquad\qquad\quad\;\; +\: {Y_{\nu_{ij}}}R_{{S^ \pm }}^{(2 + i)\alpha }Z_ + ^{2\beta }Z_n^{(4 + j)\eta } - {\lambda _i}R_{{S^ \pm }}^{1\alpha }Z_ + ^{2\beta }Z_n^{(4 + i)\eta },\\
&&C_L^{S_\alpha ^{-\ast} {\chi _\beta }\bar \chi _\eta ^ \circ } =   \frac{e}{{\sqrt 2 {s_{_W}}{c_{_W}}}}\Big( R{{_{{S^ \pm }}^{1\alpha\ast }} }Z_ - ^{2\beta } + R{{_{{S^ \pm }}^{(2 + i)\alpha }}^ * }Z_ - ^{(2 + i)\beta }\Big)\Big( {c_{_W}}Z_n^{2\eta } + {s_{_W}}Z_n^{1\eta }\Big) \nonumber\\
&&\qquad\qquad\quad\;\; - \frac{e}{{{s_{_W}}}}Z_ - ^{1\beta }\Big( R{{_{{S^ \pm }}^{1\alpha\ast }} }Z_n^{3\eta } + R{{_{{S^ \pm }}^{(2 + i)\alpha\ast }} }Z_n^{(7 + i)\eta }\Big) + {Y_{\nu_{ij}}}R_{{S^ \pm }}^{2\alpha }Z_ - ^{(2 + i)\beta }Z_n^{(4 + j)\eta }\nonumber\\
&&\qquad\qquad\quad\;\; +\: {Y_{{e_{ij}}}}R_{{S^ \pm }}^{(5 + j)\alpha }\Big( Z_ - ^{2\beta }Z_n^{(7 + i)\eta } - Z_ - ^{(2 + i)\beta }Z_n^{3\eta } \Big) - {\lambda _i}R_{{S^ \pm }}^{2\alpha }Z_ - ^{2\beta }Z_n^{(4 + i)\eta },\\
&&C_R^{S_\alpha ^ - \chi _\eta ^ \circ {{\bar \chi }_\beta }} = \Big( C_L^{S_\alpha ^{-\ast} {\chi _\beta }\bar \chi _\eta ^\circ } \Big)^*, \qquad\quad  C_R^{S_\alpha ^{-\ast} {\chi _\beta }\bar \chi _\eta ^ \circ } = \Big( C_L^{S_\alpha ^ - \chi _\eta ^ \circ {{\bar \chi }_\beta }}\Big)^*.
\end{eqnarray}

\section{Form factors\label{appendix-factor}}
Defined ${x_i} = {{{m_i^2}}/{{m_W^2}}}$, we can find the form factors:
\begin{eqnarray}
&&{I_1}(\textit{x}_1 , x_2 ) = \frac{1}{{16{\pi ^2}}}\Big[ \frac{{1 + \ln {x_2}}}{{({x_2} - {x_1})}} + \frac{{{x_1}\ln {x_1}}-{{x_2}\ln {x_2}}}{{{{({x_2} - {x_1})}^2}}} \Big],\\
&&{I_2}(\textit{x}_1 , x_2 ) = \frac{1}{{16{\pi ^2}}}\Big[ - \frac{{1 + \ln {x_1}}}{{({x_2} - {x_1})}} - \frac{{{x_1}\ln {x_1}}-{{x_2}\ln {x_2}}}{{{{({x_2} - {x_1})}^2}}} \Big],\\
&&{I_3}(\textit{x}_1 , x_2 ) = \frac{1}{{32{\pi ^2}}}\Big[  \frac{{3 + 2\ln {x_2}}}{{({x_2} - {x_1})}} - \frac{{2{x_2} + 4{x_2}\ln {x_2}}}{{{{({x_2} - {x_1})}^2}}} -\frac{{2x_1^2\ln {x_1}}}{{{{({x_2} - {x_1})}^3}}} + \frac{{2x_2^2\ln {x_2}}}{{{{({x_2} - {x_1})}^3}}}\Big], \\
&&{I_4}(\textit{x}_1 , x_2 ) = \frac{1}{{96{\pi ^2}}} \Big[ \frac{{11 + 6\ln {x_2}}}{{({x_2} - {x_1})}}- \frac{{15{x_2} + 18{x_2}\ln {x_2}}}{{{{({x_2} - {x_1})}^2}}} + \frac{{6x_2^2 + 18x_2^2\ln {x_2}}}{{{{({x_2} - {x_1})}^3}}}  \nonumber\\
&&\qquad\qquad\qquad\qquad + \: \frac{{6x_1^3\ln {x_1}}-{6x_2^3\ln {x_2}}}{{{{({x_2} - {x_1})}^4}}}  \Big].\
\end{eqnarray}

\end{document}